\documentclass[english,superscriptaddress, prb, preprint]{revtex4-1}
\usepackage[T1]{fontenc}
\usepackage[latin9]{inputenc}
\usepackage{geometry}
\usepackage{amstext}
\usepackage{amssymb}
\usepackage{graphicx}
\usepackage{wasysym}
\usepackage{esint}

\makeatletter
\usepackage[hidelinks]{hyperref}
\usepackage{xcolor}
\hypersetup{
    colorlinks,
    linkcolor={blue!80!black},
    citecolor={blue!80!black},
    urlcolor={blue!80!black}
}

\@ifundefined{showcaptionsetup}{}{%
 \PassOptionsToPackage{caption=false}{subfig}}
\usepackage{subfig}
\makeatother

\usepackage{babel}
\begin{document}

\title{Three-dimensional Modeling of Vacuum Field Emission Nanotriodes}

\author{M. S. Khalifa}

\affiliation{Center for Nanotechnology, Zewail City of Science and Technology,
Giza 12588, Egypt}

\affiliation{Engineering Mathematics and Physics Department, Faculty of Engineering,
Cairo University, Giza 12613, Egypt}

\author{A. H. Badawi}

\email[Corresponding author; ]{abadawi@zewailcity.edu.eg}

\selectlanguage{english}%

\affiliation{Center for Nanotechnology, Zewail City of Science and Technology,
Giza 12588, Egypt}

\author{T. A. Ali}

\affiliation{Center for Nanotechnology, Zewail City of Science and Technology,
Giza 12588, Egypt}

\affiliation{Engineering Mathematics and Physics Department, Faculty of Engineering,
Cairo University, Giza 12613, Egypt}

\author{N. H. Rafat}

\affiliation{Engineering Mathematics and Physics Department, Faculty of Engineering,
Cairo University, Giza 12613, Egypt}

\author{A. A. Abouelsaood}

\affiliation{Engineering Mathematics and Physics Department, Faculty of Engineering,
Cairo University, Giza 12613, Egypt}
\begin{abstract}
Vacuum nanodevices are devices that the electron transport through
them is based on electron field emission from a nano-eimtter to another
opposite electrode through a vacuum channel. Geometrically asymmetric
metal-vacuum-metal structures were demonstrated to have energy conversion
ability for electromagnetic waves in the optical range. Combining
the ability of these structures to convert optical signals into rectified
current and the ability of vacuum nanotriodes to control the field
emission current can allow direct processing on converted optical
signals using a single device. In this paper, a three-dimensional
quantum-mechanical method, rather than the approximate Fowler-Nordheim
theory, is used for modeling the field emission process in vertical-type
vacuum nanotriodes consisting of an emitter, a collector and a gate.
The electron transport through the device is computed using a transfer-matrix
technique. The potentials of vacuum nanotriodes in the current rectification
and modulation are investigated at low voltages. The effects of varying
the structure geometrical parameters on the rectified current are
also studied. The obtained results show that a great enhancement in
the rectification properties is achievable when the gate and the collector
are connected through a DC source. It is also demonstrated that a
small variation in the gate voltage can be used either to modulate
the rectified current or to switch the device into a resonant tunneling
diode.
\end{abstract}
\maketitle
\renewcommand\[{\begin{equation}} \renewcommand\]{\end{equation}}

\global\long\def\EquationCap{\text{Eq. }}

\global\long\def\Equation{\text{Eq. }}

\global\long\def\Equations{\text{Eqs. }}

\global\long\def\FigureCap{\text{Fig. }}

\global\long\def\Figure{\text{Fig. }}

\global\long\def\Figures{\text{Figs. }}

\global\long\def\FiguresCap{\text{Figs. }}

\global\long\def\SectionCap{\text{Section }}

\global\long\def\Section{\text{section }}

\global\long\def\Reference{\text{Ref. }}

\global\long\def\References{\text{Refs. }}

\section{Introduction}

The rapid progress in nanofabrication during the past several years
allowed for realizing nanostructures with minimum features of a few
nanometers. This advancement, along with the growing need for high-frequency
electronics, initiated the research in the field of vacuum nanoelectronics.
Unlike semiconductor devices, vacuum nanoelectronic devices do not
suffer from the limitation of electron velocity saturation as they
depend on ballistic electron transport, which allows operating at
higher frequencies. Also, they have greater thermal tolerance and
exhibit higher robustness against high levels of radiation, which
make them better candidates than semiconductor devices in extreme
environments such as military and space applications. On the other
hand, the active research in optical rectennas (rectifying nano-antennas)
has recently exhibited promising results in energy conversion and
current rectification. Combining the potentials of vacuum nanodevices
with optical rectennas allows for fast, local modulation of the output
current of rectennas. As a result, energy conversion, current rectification
and processing functions can all be implemented by a single device.
This opens the area to a wide field of applications in optical computing
and communication technologies, where fast logic operations can be
executed directly by rectennas.

In principle, vacuum nanoelectronic devices depend in their operation
on field emission (FE) from nanotips supported on cathodes. The emitted
electrons are usually collected at an opposite flat electrode (the
anode or collector), and a gate may be added in the electrons path
to control the magnitude of the emission current. Optical rectennas
as well depend in their rectification behavior on asymmetric electrons
emission (or tunneling) in asymmetric metal-insulator-metal (MIM)
junctions. In this work we are particularly interested in vertical
(Spindt-type) FE triodes \cite{Spindt1968}. It was demonstrated that
reducing the dimensions of the different parameters of such devices
(tip radius, gate aperture diameter, and gap distance) is a key factor
in enhancing their performance in terms of the current density, the
applied voltages and the cutoff frequency \cite{Nguyen1989,Chen2012}.

Many attempts have been made in the process of miniaturizing these
devices. In 1989, Brodie \cite{Brodie1989} discussed the physics
governing FE from a conic tip to an integrated collector electrode
at a separation of distance $0.5\text{ }\mbox{\ensuremath{\mu}m}$,
with gate aperture of radius $0.5\text{ }\mbox{\ensuremath{\mu}m}$.
In 2000, Driskill-Smith et al. \cite{Driskill-Smith2000} fabricated
long nanopillars of radius $1\text{ }\mbox{nm}$ in a vacuum nano-chamber
with dimensions of about $0.1\text{ }\mbox{\ensuremath{\mu}m}$. More
recently, materials such as CNTs and Si nanowires have been used in
fabrication as the emitting nanotips, for their high aspect ratio
and relatively low operating voltages, with self-aligning gate around
them \cite{Pflug2001,Guillorn2001,Gangloff2004,Wong2005,Ulisse2012}.
This allowed minimizing the gate aperture down to $45\text{ }\mbox{nm}$
in radius \cite{Chen2012}.

In $\References$ \onlinecite{Chen2012,Brodie1989,Driskill-Smith2000,Pflug2001,Guillorn2001,Gangloff2004,Wong2005,Ulisse2012},
the authors described the FE process based on Fowler-Nordheim (FN)
theory, in which the current density is expressed as a function of
the electric field at the emitter surface. In the original FN theory,
the emitting surface is assumed to be planar, and hence a one-dimensional
problem is considered. This assumption is accurate as long as the
emitter radius is much greater than the potential barrier, where the
electric field can be considered uniform along the emitting surface.
However, when the emitter radius is comparable to or smaller than
the barrier width, the one-dimensional solution is no longer valid,
and tunneling through a three-dimensional potential barrier should
be considered instead. Although many correcting factors were introduced
to FN basic equation for considering the emitter geometry and size
\cite{JunHe1991,Cutler1993,Jensen1995,Nicolaescu2001}, they were
demonstrated to be inaccurate when applied to sharp emitters with
radii $\apprle10\text{ }\mbox{nm}$ \cite{Cutler1993}. Another effective
classical model, so called Quantum Corrected Model (QCM), was developed
more recently for modeling electron tunneling through plasmonic nanogaps
\cite{Esteban2012}. However, this model also assumes a potential
barrier that is much smaller than the radius of the emitting surface,
which is the typical case in plasmonic systems. For describing the
behavior of vacuum nanotriodes without implying modeling restrictions,
one can use either the Green-Function method \cite{Lucas1988,Doyen1993}
or the Transfer-Matrix method \cite{Mayer1997,Mayer1999}. In each
of these methods, Schr{\"o}dinger's equation is solved in three dimensions.
In this work we follow the transfer-matrix method, which is more suitable
in terms of memory storage when dealing with emitters with a few nanometers
in height \cite{Mayer1999}.

In this paper, the model proposed by Mayer et al. in $\Reference$\onlinecite{Mayer2008}
for a high-frequency rectifier is extended by introducing a metallic
gate to the structure for controlling the current flow through the
device, so that the device resembles Spindt-type vacuum field emission
triode. We mainly investigate the effect of the gate on the I-V characteristics
of the device. We also study the effect of the gate voltage on the
rectification properties of the device in the limit of quasi-static
bias. In $\Section$\ref{sec:Methodology}, we state the assumptions
made in the model and present the method we follow in computing the
potential energy distribution and the emission current, with emphasizing
the modifications we introduced for considering the gate effect. The
behavior of the device is then studied in $\Section$\ref{sec:Calculations-and-Results}.
First, we present and discuss the modifications introduced to the
potential barrier in the device due to the effect of the gate. Next,
we investigate the possibility of improving the current rectification
and the output power of the device in different situations for the
gate potential. Then, we investigate the effect of the thickness,
the height and the aperture diameter of the gate on the output current
of the device. Finally, we investigate how changing the height and
the diameter of the emitting tip can be exploited for improving the
behavior of the device.

\begin{figure}
\noindent \begin{centering}
\includegraphics[scale=0.35]{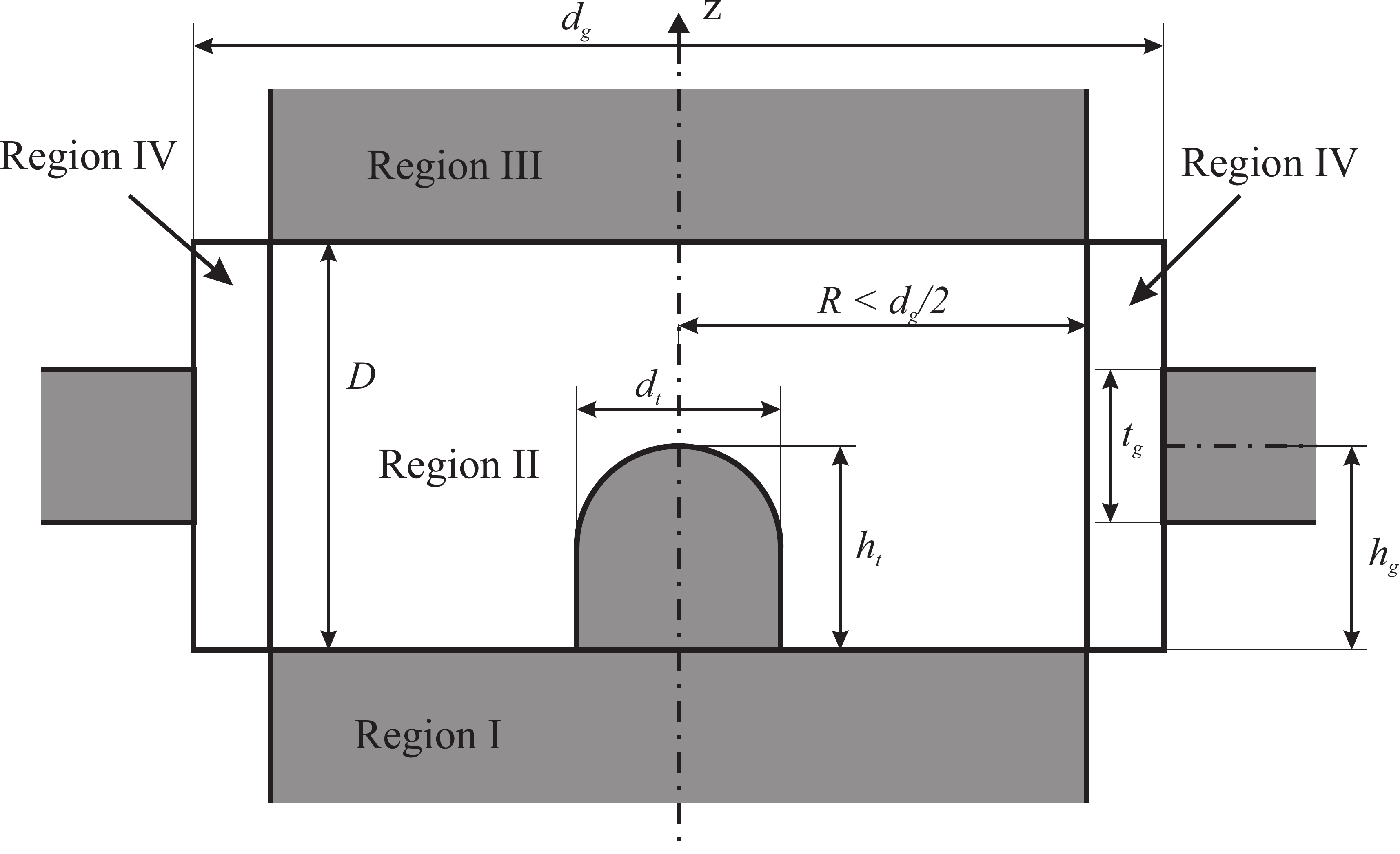}
\par\end{centering}

\caption{Schematic representation for the proposed structure. The rectangles
at the sides represent the gate disc which surrounds the rectifier
circularly, and are extended in the radial direction.\label{fig:The-proposed-structure}}

\end{figure}

\section{Methodology\label{sec:Methodology}}

\subsection{Preliminaries\label{sub:Preliminaries}}

The structure we study in this paper is shown in $\Figure$\ref{fig:The-proposed-structure}.
It consists of two metallic parallel flat planes separated by a distance
$D$, with the lower one supporting a metallic cylindrical nanotip
of a hemispherical end whose height and diameter are $h_{t}$ and
$d_{t}$, respectively. Similar structures with geometrical asymmetry
were proven to exhibit rectification properties both theoretically
\cite{Mayer2008,Miskovsky1979} and experimentally \cite{Kuk1990,Bragas1998,Tu2006,Dagenais2010,Ward2010}.
A gate electrode, represented by an infinite horizontal metallic disc
of thickness $t_{g}$, is set at height $h_{g}$ above the lower metal,
and contains a circular aperture of diameter $d_{g}$ concentric with
the tip. The rest of the space between the two surfaces is assumed
to be vacuum.

We consider the metallic planes as two long leads (regions I and III)
of radius $R$. The leads are assumed to be perfect conducting metals,
that is, electrons inside them have a uniform potential energy. The
objective of this section is to study the quantum transport between
region I and region III through some quantum device (region II) representing
the volume enclosed by the horizontal planes $z=0$ and $z=D$, and
the lateral surface defined by $\rho=R$. The vacuum cylindrical shell
between the end of the leads at $\rho=R$ and the start of the gate
at $\rho=d_{g}/2$ (region IV) is considered outside the device. Electrons
are assumed to be confined in the leads and the device within the
cylindrical space of radius $R$, that is, the leakage current in
the gate is neglected in this model and left for future investigations.
This approximation is acceptable in the light of the results obtained
by Driskill-Smith et al. \cite{Driskill-Smith2000}, where the calculations
of the electrons trajectories showed that all electrons emitted from
the tip are collected at the anode for all anode voltages except at
zero volt.

We note here that $R$ should not exceed the gate aperture radius,
$d_{g}/2$, for two physical reasons. The first reason is that stacking
three metallic layers with two vacuum barriers in between would result
in resonant tunneling between the leads at the regions where the
three layers overlap \cite{Ricco1984}. This will then overwhelm the
rectification behavior of the device as well as the current control
by the gate. The second reason is to reduce the gate-cathode and the
gate-anode capacitances that would decrease the cutoff frequency of
the device \cite{Ulisse2012}. In order to treat this situation in
the analysis, we solve Poisson's equation in regions II and IV with
taking the gate effect (both the potential on the gate as well as
its metallic effect regardless of the metal type) as a boundary condition
on the lateral edge of region IV at $\rho=d_{g}/2$ and at $z$ between
$h_{g}-t_{g}/2$ and $h_{g}+t_{g}/2$. Then we use the obtained value
of the potential in region II only, where electrons are assumed to
be localized, to solve Schr{\"o}dinger's equation. The material properties
of the gate, namely the workfunction and the Fermi energy, are not
important in our calculations, because the leakage current is essentially
assumed to be neglected, and the gate itself is taken outside the
solution region which is limited by $\rho=R$.

When the structure is exposed to an external electric field directed
along the axis of the device, $z$-direction, a potential difference
is induced between the two leads with a magnitude that depends on
the intensity of the field and the length of the device, $D$ in $\Figure$\ref{fig:The-proposed-structure},
where $\Delta V=-ED$. This assumption is valid when considering an
electrostatic field or a quasi-static electric field that could be
carried by an incident electromagnetic wave with a relatively low
frequency. The quasi-static limit here applies for waves whose periods
are much longer than the average time an electron would take to transport
through the device. For the structure parameters we consider in $\Section$\ref{sec:Calculations-and-Results},
the quasi-static limit is valid for small frequencies compared to
$1000\text{ }\mbox{THz}$ \cite{Mayer2008}. In all the upcoming analysis,
we account for the external potential difference $\Delta V$ between
the collector (upper lead) and the emitter (lower lead) by taking
its value as a voltage applied to the collector; $V_{c}=\Delta V$,
while the emitter is kept grounded. Similarly, we take the voltage
applied to the gate $V_{g}$ referred to the emitter.

\subsection{Potential Energy Distribution\label{sub:Potential-Energy-Distribution}}

The first step in the analysis is obtaining the electron potential
energy distribution in each of the four regions in $\Figure$\ref{fig:The-proposed-structure}
in order to solve Schr{\"o}dinger's equation in regions I, II, and III
accordingly. Metallic leads are considered in regions I and III with
a work function $W$ and a Fermi energy $E_{f}$. The electron potential
energy in these regions are then $U^{I}=U^{III}=-(W+E_{f})$. When
external voltage $V_{c}$ is applied to the upper lead, another term
$-eV_{c}$ is added to the potential energy in region III, so that
$U^{III}=-(W+E_{f})-eV_{c}$, where $e$ is the magnitude of the electron
charge.

Unlike the previous two regions, regions II and IV include non-uniform
potential energy distribution. Electron potential energy at any point
in the vacuum in these regions consists of two parts; $U_{bias}$
and $U_{met}$. The first part, $U_{bias}$, is the potential energy
induced from the externally applied voltages on the leads and the
gate. The electric potential distribution due to this bias, $V_{bias}$,
can be obtained by solving Poisson's equation in the volume of regions
II and IV, taking the following boundary conditions. In region II
we take ground potential on the lower surface and the tip, and a constant
potential $V_{c}$ on the upper surface. In region IV, we take a zero
charge boundary condition on the lower and the upper surfaces. On
the lateral surface at $\rho=d_{g}/2$ there are three domains; a)
$0<z<h_{g}-t_{g}/2$, b) $h_{g}-t_{g}/2<z<h_{g}+t_{g}/2$, and c)
$h_{g}+t_{g}/2<z<D$. The zero charge boundary condition is taken
along the first and the third domains, while a constant potential
$V_{g}$ is taken along the second domain which represents the gate
surface. The zero charge boundary condition indicates that the normal
component of the electric field is zero at the boundary, i.e. the
electric potential is constant along the normal direction. The second
part, $U_{met}$, is the self-induced potential energy by the tunneling
electron during its transport through the device. This part represents
the potential energy of the electron due to the accumulated charges
on the metallic surfaces of the leads, the tip and the gate. In this
model, the electron image is taken on each metal surface individually
as a first approximation. This enables calculating this part of the
potential by solving Poisson's equation numerically \cite{Laloyaux1993,Mayer2005},
with the boundary condition {\small{}$V_{met}(\mathbf{r_{b}})=\frac{1}{4\pi\epsilon_{0}}\frac{e}{|\mathbf{r_{b}}-\mathbf{r_{e}}|}$}
on all the metallic surfaces, including the gate, where $\mathbf{r}_{e}$
is the position of the electron and $\mathbf{r_{b}}$ is a point on
the conductor at which the boundary potential is calculated. The zero
charge boundary condition is taken here again on the non-metallic
boundaries in region IV. The potential energy of the electron due
to this electric potential is $U_{met}^{II}(\mathbf{r_{e}})=-eV_{met}(\mathbf{r}_{e})/2$,
where the factor $1/2$ arises from the fact that this potential energy
is self-induced by the electron\cite{smythe1950}. In order to avoid
the divergence of this term near the metallic surfaces, we cut all
the values lower than the potential energy inside the metals and set
them by this value \cite{Murphy1956,Modinos2001}. Since the metallic
tip is supported on the grounded lead, electrons in the tip have zero
$U_{bias}^{II}$ and constant $U_{met}^{II}=-(W+E_{f}^{tip})$, where
$E_{f}^{tip}$ is the Fermi energy of the material of the tip.

\subsection{Field Emission Current\label{sub:Field-Emission-Current}}

Now we proceed to solving Schr{\"o}dinger's equation, using the obtained
potential energy, in order to get the FE current. Since electrons
are assumed to be confined in a cylinder of radius $R$, then their
wavefunctions can be expanded in terms of complete, orthonormal eigenstates
in cylindrical coordinates as follows\cite{Mayer1997}

\[
\Psi(\rho,\phi,z)={\displaystyle \sum_{m,j}}\Phi_{mj}(z)\frac{J_{m}(k_{mj}\rho)}{\sqrt{\int_{0}^{R}\rho\left[J_{m}(k_{mj}\rho)\right]^{2}d\rho}}\frac{e^{im\phi}}{\sqrt{2\pi}}
\]
where $J_{m}$ is the $m^{th}$ order of the Bessel function of the
first kind, with $m$ integer, and $k_{mj}$ is the $j^{th}$ coefficient
satisfying $J_{m}^{'}(k_{mj}R)=0$, with $j$ positive integer. $\Phi_{mj}(z)$
are the coefficients of the eigenstates depending on the coordinate $z$.
In regions I and III, each lead has a constant potential energy and
is considered semi-infinite in the $z$-direction, therefore for an
electron with energy $E$ the coefficients are  {\small{}$\Phi_{mj}^{I/III}(z)=\alpha e^{\pm ik_{z,mj}^{I/III}z}$},
where {\small{}$k_{z,mj}^{I/III}=\sqrt{\frac{2m_{e}}{\hbar^{2}}\left(E-U^{I/III}\right)-k_{mj}^{2}}$}
and $\alpha$ is a normalization factor. The $\pm$ sign indicates
the propagation direction relative to $z$-axis.

To obtain the transmission probability through the quantum device
(region II), the transfer-matrix method developed in $\References$\onlinecite{Mayer1997,Mayer1999,Mayer1999a}
is followed. In this technique the potential energy in region II is
divided into two parts according to its coordinates-dependency; main
potential $U_{0}^{II}(z)$ and local perturbing potential $U_{1}^{II}(\rho,\phi,z)$,
such that $U^{II}=U_{0}^{II}(z)+U_{1}^{II}(\rho,\phi,z)$. After mathematical
manipulations, a matrix equation coupling the coefficients $\Phi_{mj}(z)$
is obtained. Next, by discretizing region II into horizontal layers
and assuming that $U^{II}$ is independent of $z$ inside each single
layer, the matrix equation can be solved for all $\Phi_{mj}(z)$ in
each layer, as illustrated in appendix A in $\Reference$\onlinecite{Mayer1997}.
This enables calculating the scattering parameters for each layer
individually. The scattering parameters of consecutive layers are
then combined iteratively, using the layer addition algorithm \cite{Pendry1994,Mayer1999},
until one gets the total scattering parameters of region II.

Let $S^{++}$ and $S^{--}$ be the forward (from region I to III)
and reverse (from region III to I) transmission matrices of the obtained
scattering parameters. Then, the upward $I^{+}$ and the downward
$I^{-}$ electron currents can be obtained from the following expressions,
respectively \cite{Mayer2008}

\begin{equation}
I^{+}=\frac{2e}{h}{\displaystyle \int_{U_{I}}^{\infty}}f_{I}(E)\left[1-f_{III}\left(E\right)\right]{\displaystyle \sum_{mj}\sum_{m'j'}\left|S_{(m',j'),(m,j)}^{++}\right|^{2}\frac{k_{z_{m',j'}}^{III}}{k_{z_{mj}}^{I}}}dE\label{eq:UpwardCurrent}
\end{equation}

\begin{equation}
I^{-}=\frac{2e}{h}{\displaystyle \int_{U_{III}}^{\infty}}f_{III}(E)\left[1-f_{I}\left(E\right)\right]{\displaystyle \sum_{mj}\sum_{m'j'}\left|S_{(m',j'),(m,j)}^{--}\right|^{2}\frac{k_{z_{m',j'}}^{I}}{k_{z_{mj}}^{III}}}dE\label{eq:DownwardCurrent}
\end{equation}
where $f_{I}(E)$ and $f_{III}(E)$ are the Fermi functions at regions
I and III, whose Fermi levels are given by $\mu_{I}=-W$
and $\mu_{III}=-W-eV_{c}$, respectively. The temperature in the Fermi functions
is taken to be $300\text{ }\mbox{K}$.

For a finite number of modes and finite matrices dimensions, a finite
number of values for the quantum numbers $m$ and $j$ should be considered.
Ideally, all the values of $m$ and $j$ satisfying the relation 

\[
k_{mj}\le\sqrt{\frac{2m_{e}}{\hbar^{2}}\left(E-min\left(U^{I},U^{III}\right)\right)}
\]
should be included. This condition ensures including all the modes
propagating in at least one of the two leads, where these are the
responsible modes for conducting current through the device. However,
in the structure we study here, the existence of the tip around the
$z$-axis makes the modes associated with small values of $|m|$ (the
modes with high probability around the center) have higher contribution
to the tunneling current than those associated with larger values
of $|m|$. This is because the Bessel functions $J_{m}$ have higher
values near the center for smaller values of $|m|$, \cite{Mayer2008a}.
This allows us to consider a) all modes with $|m|\le m_{max}$, where
$m_{max}$ is as high as necessary for reaching convergence, and b)
all values of $j$ satisfying the above condition on $k_{mj}$ for
the associated $m$. In this work we take $m_{max}=4$.

\section{Results and Discussion\label{sec:Calculations-and-Results}}

In this section, we aim at exploring the potentials of the FE nanotriode
for current rectification and modulation at low voltages. In particular,
we investigate the effect of the gate in modulating the behavior of
the structure and whether it can be exploited for enhancing the current
rectification and the mean output power of the device. We also study
how the geometrical parameters of the device affect its performance
and how they can be optimized for current modulation through the gate
voltage.

\subsection{Potential Barrier Modulation\label{sub:Barrier-Modulation}}

In this section, we study the effect of the gate on the shape of the
potential barrier. We consider cylindrical leads with radius $R=2\text{ }\mbox{nm}$,
separated by distance $D=2\text{ }\mbox{nm}$, as shown in $\Figure$\ref{fig:The-proposed-structure},
and made of tungsten whose work function and Fermi energy are $4.5$
and $19.1\text{ }\mbox{eV}$, respectively. In region II a nanotip
of height $h_{t}=1\text{ }\mbox{nm}$ and diameter $d_{t}=1\text{ }\mbox{nm}$
is set on the lower lead, and made of tungsten as well. The rest of
region II is assumed to be vacuum. Outside region II we consider a
gate of thickness $t_{g}=1\text{ }\mbox{nm}$ set at height $h_{g}=1\text{ }\mbox{nm}$
above the lower lead, and aperture diameter $d_{g}=4.1\mbox{\text{ }}\mbox{nm}$.
The value of the gate aperture diameter is chosen to be greater than
the leads diameter in order to avoid the resonant tunneling. Both
the thickness and the height of the gate are chosen such that the
gate is centered vertically at the end of the tip. Both the collector
bias $V_{c}$ and the gate bias $V_{g}$ are taken to be $1\text{ }\mbox{V}$.
As we mentioned before in $\Section$\ref{sub:Preliminaries}, the
material properties of the gate are not important in calculations,
therefore all the results in this section are applicable to a gate
of any metal that can be considered a perfect conductor.

According to the model assumptions in $\Section$\ref{sec:Methodology},
the following results and discussion are applicable only to the limit
of quasi-static fields. This limit depends on the cutoff frequency
of the device, which is given by half of the reciprocal of the average
time taken by an electron to travel between the emitter and the collector;
the traversal time \cite{Bttiker1982}. In the classically forbidden
region, an electron is assumed to be traveling at the Fermi velocity
\cite{Nguyen1989,Sullivan1989}, while outside the barrier the electron
propagates classically with a velocity that is proportional to its
wavevector. Having the velocity and the gap distance between the emitter
and the collector, the traversal time of the electrons can be calculated.
For the structure parameters mentioned above, the traversal time takes
a value of $0.5\text{ }\mbox{fs}$ \cite{Mayer2008}, and the cutoff
frequency is, therefore, $1000\text{ }\mbox{THz}$, which corresponds
to electromagnetic wavelength of about $300\text{ }\mbox{nm}$, that
is in the ultraviolet range. The quasi-static limit assumption is
valid for frequencies significantly lower than this value. In case
of oscillating fields with frequencies close to the cutoff frequency,
photon absorption and emission should be included in the model \cite{Mayer2000a,Miskovsky1994}.
This problem will be treated in future work, including the gate voltage
oscillation.

\begin{figure}
\begin{centering}
\includegraphics[scale=0.65]{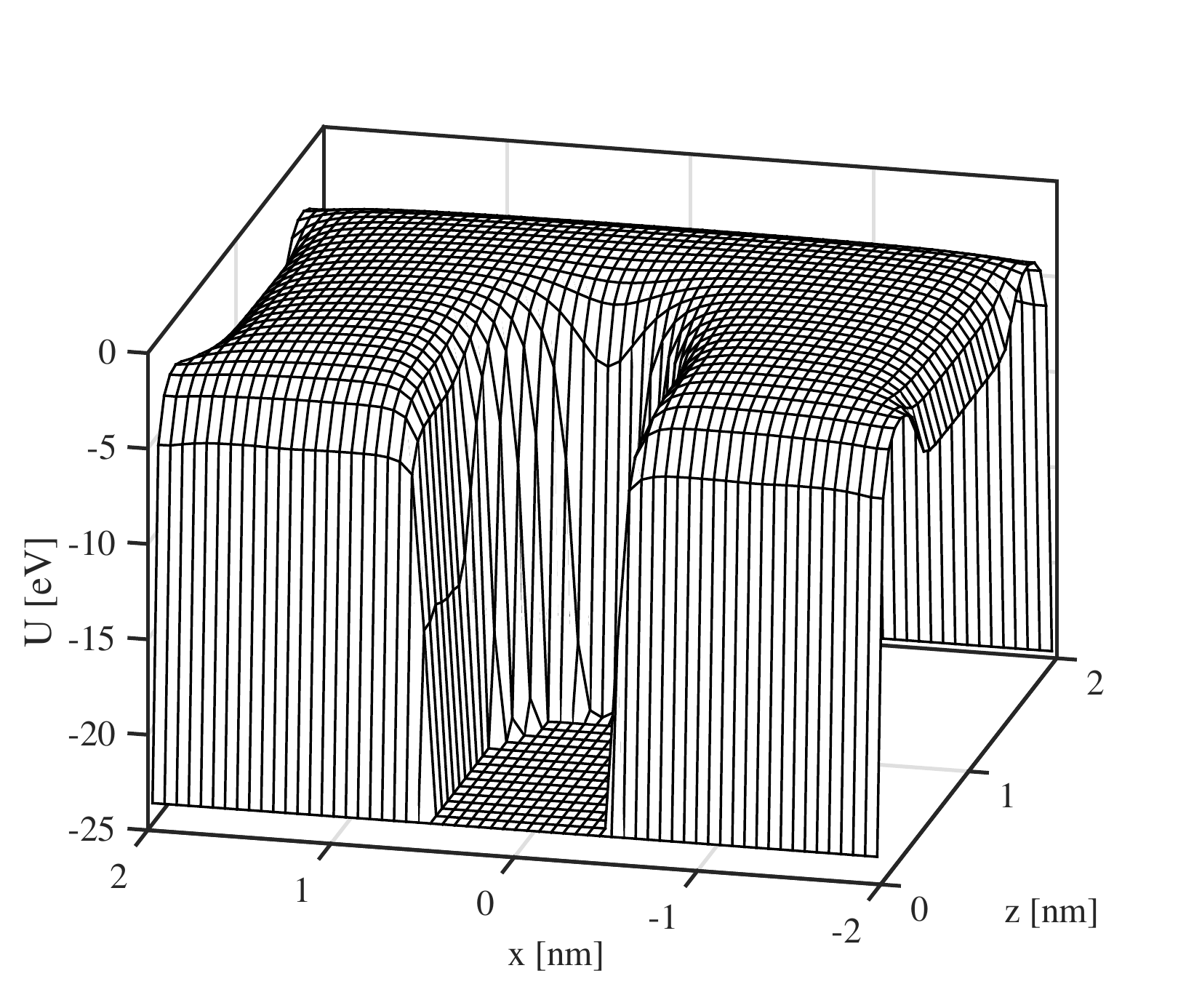}
\par\end{centering}

\caption{Potential energy distribution in $xz$-plane at equal collector and
gate voltages of $1\text{ }\mbox{V}$. The gate thickness $t_{g}$,
height $h_{g}$ and aperture diameter $d_{g}$ are $1$, $1$ and
$4.1\text{ }\mbox{nm}$, respectively.\label{fig:Potential-Energy-Distribution}}

\end{figure}

\begin{figure}
\subfloat[\label{fig:Potential-Barrier-1-a}]{\begin{centering}
\includegraphics[scale=0.65]{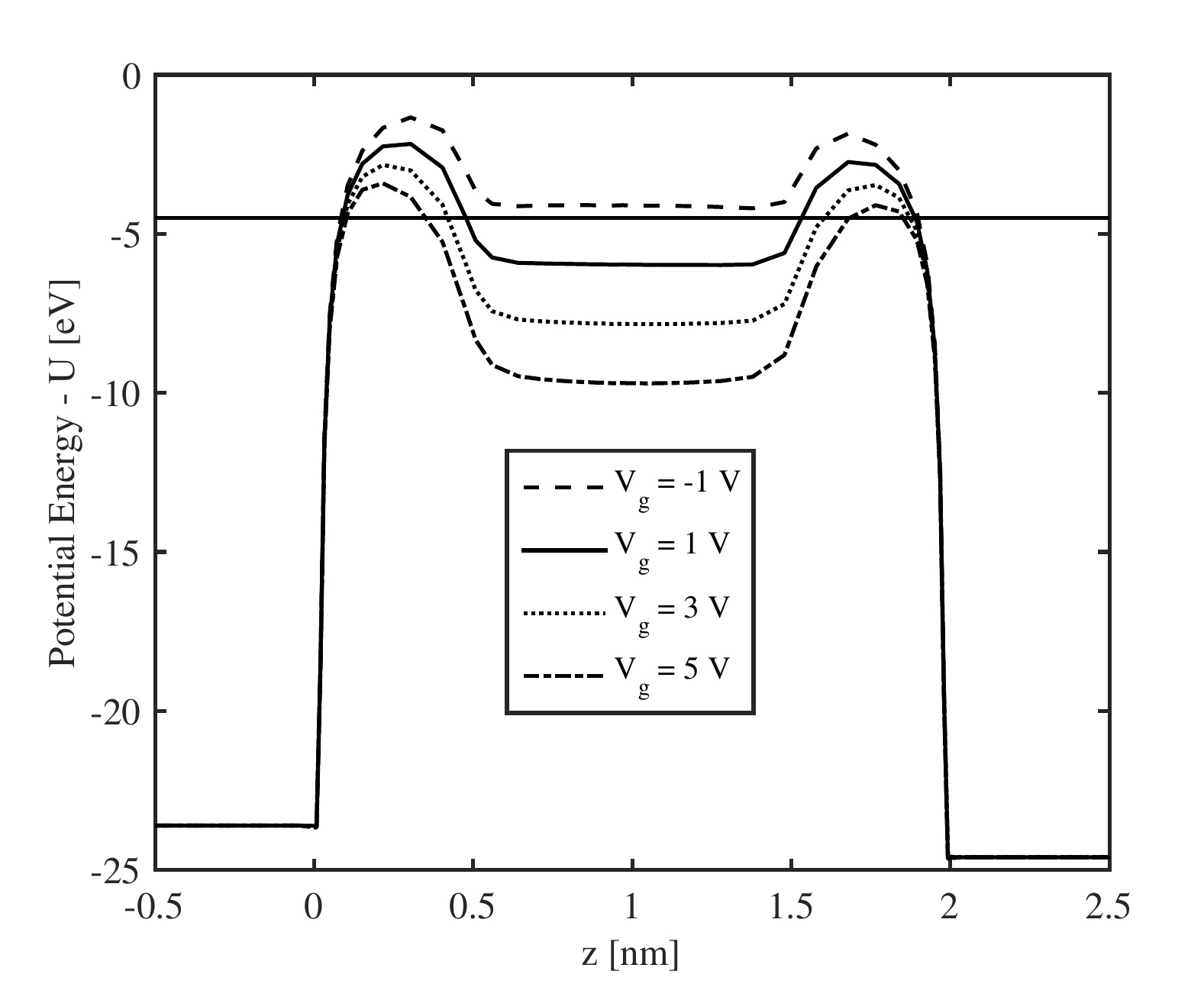}
\par\end{centering}

}

\subfloat[\label{fig:Potential-Barrier-1-b}]{\begin{centering}
\includegraphics[scale=0.65]{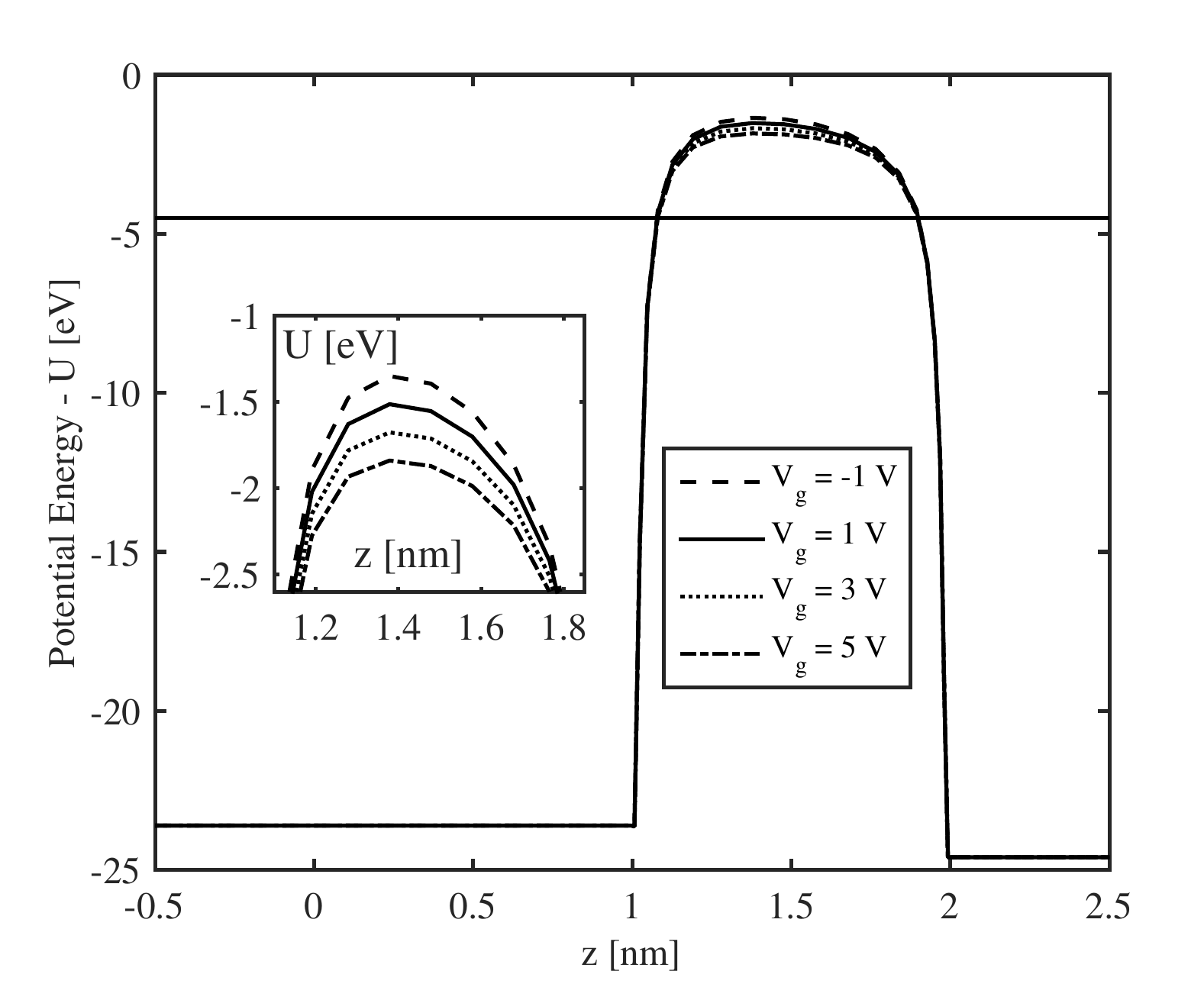}
\par\end{centering}

}

\caption{Potential energy along $z$-axis at (a) $\rho=2\text{ }\mbox{nm}$
and (b) $\rho=0$ when $V_{c}=1\text{ }\mbox{V}$ and $V_{g}=-1$,
$1$, $3$ and $5\text{ }\mbox{V}$. The peak of the barrier at $\rho=0$
is magnified in the inset in (b) in order to show small changes in
the barrier height at different gate voltages. The gate thickness
$t_{g}$, height $h_{g}$ and aperture diameter $d_{g}$ are $1$,
$1$ and $4.1\text{ }\mbox{nm}$, respectively. The horizontal black
solid lines indicate the Fermi level inside the emitter.\label{fig:Potential-Barrier-1}}

\end{figure}

Due to the axial symmetry of the structure about the $z$-axis, electrons
potential energy and hence their wavefunctions are independent of
$\phi$. Thus, it is sufficient to calculate the potential energy
at a single plane defined by a certain $\phi$. This allows calculating
the bias potential $V_{bias}$ at a single $\rho z$-plane by solving
Poisson's equation in two dimensions. In contrast, when calculating
the image potential $V_{met}$, Poisson's equation is solved in three
dimensions because the charge accumulation on the metallic surfaces
is not axially symmetric for off-axis electron position. The axial
symmetry, however, can be exploited in calculating $V_{met}$ by considering
the electron positions in a definite $\rho z$-plane (e.g. the $xz$-plane).
Using COMSOL simulation tool, we calculate the two terms of the potential
energy in region II; $U_{bias}^{II}$ and $U_{met}^{II}$, as illustrated
in $\Section$\ref{sub:Potential-Energy-Distribution}. The total
potential energy distribution in region II, $U_{bias}^{II}+U_{met}^{II}$,
is shown in $\Figure$\ref{fig:Potential-Energy-Distribution}. The
corresponding potential barriers for electrons tunneling through the
device at the lateral boundary ($\rho=2\text{ }\mbox{nm}$) and at
the center ($\rho=0$) along the $z$-direction are drawn in solid
lines in $\Figures$\ref{fig:Potential-Barrier-1-a} and \ref{fig:Potential-Barrier-1-b},
respectively. A potential well appears at $\rho=2\text{ }\mbox{nm}$.
This fast decrease in the potential energy distribution near the lateral
boundary is mainly due to the image potential of the electron on the
gate surface. The potential well may lead to resonant tunneling, specifically
if one or more of the resonant energies are around the Fermi level
of the emitter. If the minimum resonant energy is, however, significantly
larger than the Fermi level of the emitter, no considerable resonant
tunneling occurs because the electrons occupation for resonant energies
is almost zero. The resonant tunneling effect, as mentioned before,
is undesirable because it leads to high level of conduction in both
directions which reduces the rectification effect of the tip.

It is interesting to see how changing the gate voltage would modulate
the shape of the barrier and the well. $\FigureCap$\ref{fig:Potential-Barrier-1}
shows the potential energy along $z$-axis for four different values
of $V_{g}$; $-1$, $1$, $3$ and $5\text{ }\mbox{V}$, at both $\rho=2\text{ }\mbox{nm}$
(a) and $\rho=0$ (b). These voltages are chosen to show the barrier
modulation around the Fermi level near the edge. As the gate voltage
increases the depth of the well at the edge increases, while the potential
barrier at the center slightly decreases. The increasing depth of
the well results in decreasing the resonant energy levels down to
the vicinity of the Fermi level. Although the resonant tunneling effect
is not favorable from the rectification point of view, the dependence
of the well depth and the resonant levels on the gate voltage may
be useful from another perspective, where the device can operate as
a resonant tunneling diode with controllable potential well. The more
interesting part is that with only changing the gate voltage the device
can be switched between the rectification mode and the resonant tunneling
mode. As shown in $\Figure$\ref{fig:Potential-Barrier-1-a} at $V_{g}=-1\text{ }\mbox{V}$,
the Fermi level is totally buried under the potential energy along
the $z$-axis, which means that there is no chance for resonant tunneling
to occur. However, a more comprehensive study for the device in the
case of resonant tunneling is still needed, which is beyond the scope
of this paper.

The peaks of the potential energy curves at $\rho=0$ are magnified
in the inset in $\Figure$\ref{fig:Potential-Barrier-1-b} in order
to show the small changes in the barrier height and width at different
gate voltages. As the gate voltage increases, the barrier gets slightly
lower and narrower. These small changes indicate that the effect of
the gate significantly decays as we go from the lateral boundary inwards
until it is minimally pronounced at the center.

\begin{figure}
\begin{centering}
\includegraphics[scale=0.65]{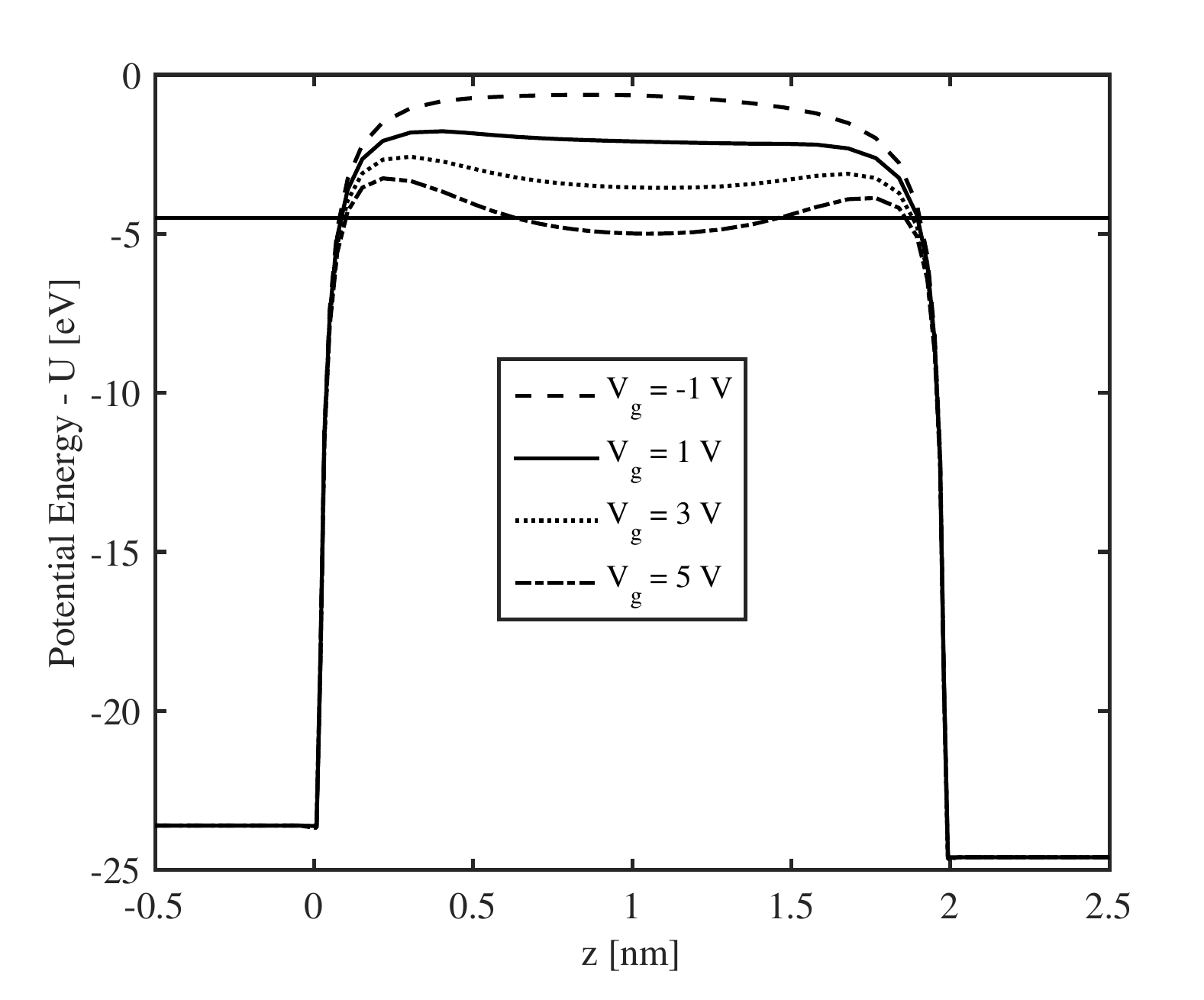}
\par\end{centering}

\caption{Potential energy along $z$-axis at $\rho=2\text{ }\mbox{nm}$ when
$V_{c}=1\text{ }\mbox{V}$ and $V_{g}=-1$, $1$, $3$ and $5\text{ }\mbox{V}$.
The gate thickness $t_{g}$, height $h_{g}$ and aperture diameter
$d_{g}$ are $1$, $1$ and $4.5\text{ }\mbox{nm}$, respectively.
The horizontal black solid line indicates the Fermi level inside the
emitter.\label{fig:Potential-Barrier-2}}

\end{figure}

In order to investigate the effect of the gate aperture diameter $d_{g}$,
we repeated the above calculations for $d_{g}=4.5\text{ }\mbox{nm}$.
The potential energies at the lateral boundary are depicted in $\Figure$\ref{fig:Potential-Barrier-2}.
The image potential on the gate surface almost vanished when the gate
diameter increased from $4.1$ to $4.5\text{ }\mbox{nm}$. This result
caused the potential well to totally disappear at gate voltages lower
than $5\text{ }\mbox{V}$, which means that the gate voltage can reach
higher values without the device encountering resonant tunneling.
Regarding the potential barrier at $\rho=0$, there is no significant
difference due to increasing $d_{g}$ except for a slight decrease
in the change of the potential barrier with changing the gate voltage.

\begin{figure}
\begin{centering}
\includegraphics[scale=0.65]{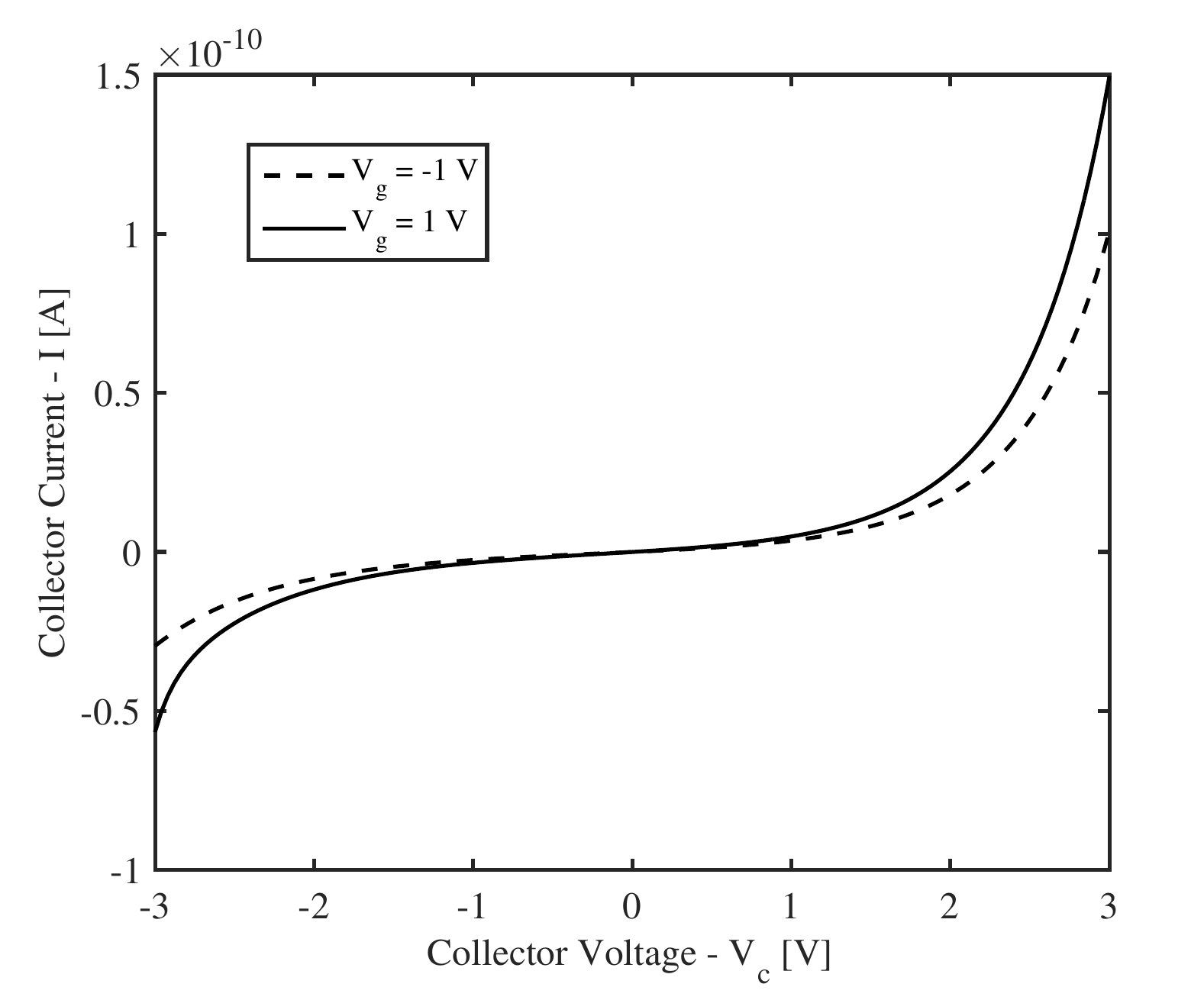}
\par\end{centering}

\caption{Current-voltage characteristics of the triode at $V_{g}=\pm1\text{ }\mbox{V}$.
The gate thickness $t_{g}$, height $h_{g}$ and aperture diameter
$d_{g}$ are $1$, $1$ and $4.5\text{ }\mbox{nm}$, respectively.\label{fig:Current-voltage-characteristics}}

\end{figure}

The current-voltage characteristics of a device of the previous parameters
at the cases of $V_{g}=-1$ and $1\text{ }\mbox{V}$ are shown in
$\Figure$\ref{fig:Current-voltage-characteristics}. The results
show the current modulation effect associated with the barrier modulation,
while keeping the rectification nature of the device. This demonstrates
the possibility of implementing electronic processes directly on the
rectified current through applying a gate voltage of a few volts in
magnitude. Both the rectification and the modulation abilities of
the device are discussed in more detail in the following two sections.

Finally, we note that the width and the position of the well along
the $z$-axis at $\rho=2\text{ }\mbox{nm}$ can be controlled by the
thickness and the height of the gate, respectively. Similar to the
dependence of the depth of the well on the gate voltage, the dependence
of the width of the well on the gate thickness fades away as the gate
aperture diameter increases. A more detailed investigation for the
effect of the gate thickness and height is presented in $\Section$\ref{sub:Current-Modulation}.

\subsection{Gate Effect on Current Rectification and Modulation\label{sub:Current-Rectification}}

In this section, we investigate how the existence of the gate modifies
the rectification properties of the device. We are interested here
in calculating both the forward and backward currents. For a positive
external bias $V_{c}$, the forward current is given by $I_{fwd}=I^{+}-I^{-}$,
while for a negative external bias, the backward current is given
by $I_{bwd}=I^{-}-I^{+}$, where $I^{+}$ and $I^{-}$ are the upward
and downward currents whose expressions were given in $\Equations$(\ref{eq:UpwardCurrent})
and (\ref{eq:DownwardCurrent}), respectively. The ability of the
device to deliver a higher current in the forward bias than the backward
when subject to the same absolute value $V_{c}$ is measured by the
rectification ratio $Rect=I_{fwd}/I_{bwd}$. If an oscillating field
of a frequency significantly lower than the cutoff frequency is incident
on the device, an oscillating potential difference of magnitude $V_{c}$
is induced between the collector and the emitter. The device is then
supposed to deliver an asymmetric oscillating current between $I_{fwd}$
and $-I_{bwd}$. This indicates that when operating as a power source,
the device is capable of delivering an output DC power given by $\left\langle P\right\rangle =\frac{1}{2}V_{c}(I_{fwd}-I_{bwd})=\frac{1}{2}V_{c}I_{fwd}(1-1/Rect)$
\cite{Mayer2008}. From this expression, we see that the output power
can be optimized by increasing the forward current and the rectification
ratio.

We now consider three cases for the gate connection; a) the gate is
connected through a DC source $V_{DC}$ to the emitter, b) the gate
is connected through a DC source $V_{DC}$ to the collector, and c)
the gate has a floating potential. In the first two cases we consider
three values for $V_{DC}$: $2\text{ }\mbox{V}$, $0$ and $-2\text{ }\mbox{V}$.
This corresponds to $V_{g}=2\text{ }\mbox{V}$, $0$ and $-2\text{ }\mbox{V}$
in the first case, and $V_{g}=V_{c}+2\text{ }\mbox{V}$, $V_{c}$
and $V_{c}-2\text{ }\mbox{V}$ in the second case.
Since we are targeting low voltages, the two values $2$ and $-2\text{ }\mbox{V}$
are taken as the extremes of the spanning range of $V_{DC}$. In the
following results we consider the same geometrical parameters mentioned
in $\Section$\ref{sub:Barrier-Modulation} with $d_{g}=4.5\text{ }\mbox{nm}$.
Since the gate electrode is centered at the midpoint between the collector
and the emitter, therefore in the third case the floating gate voltage
is driven to the mid-value between their voltages ($V_{g}\approx V_{c}/2$).

\begin{figure}
\subfloat[\label{fig:Emitter-Connection}]{\begin{centering}
\includegraphics[scale=0.65]{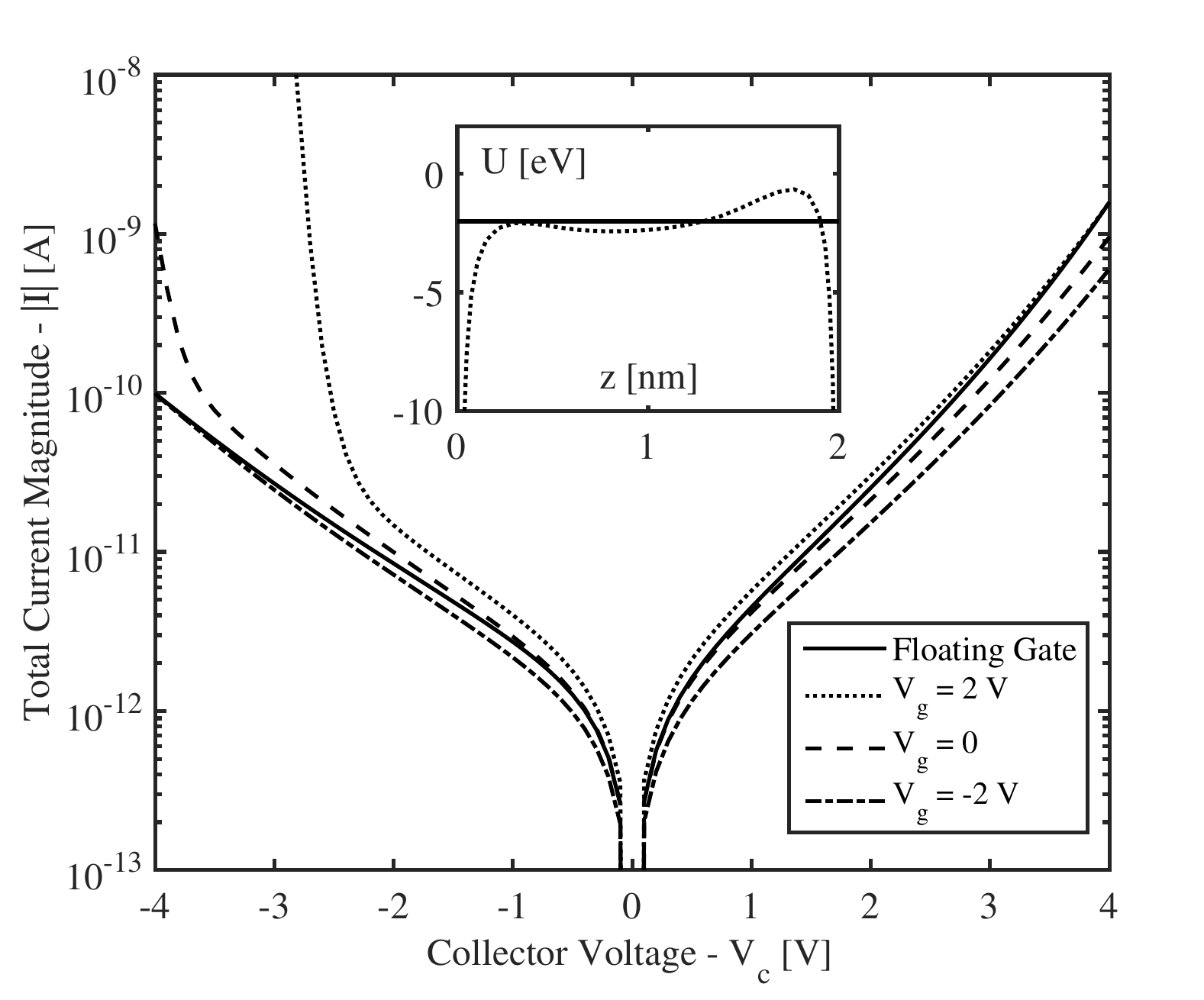}
\par\end{centering}

}

\subfloat[\label{fig:Collector-Connection}]{\begin{centering}
\includegraphics[scale=0.65]{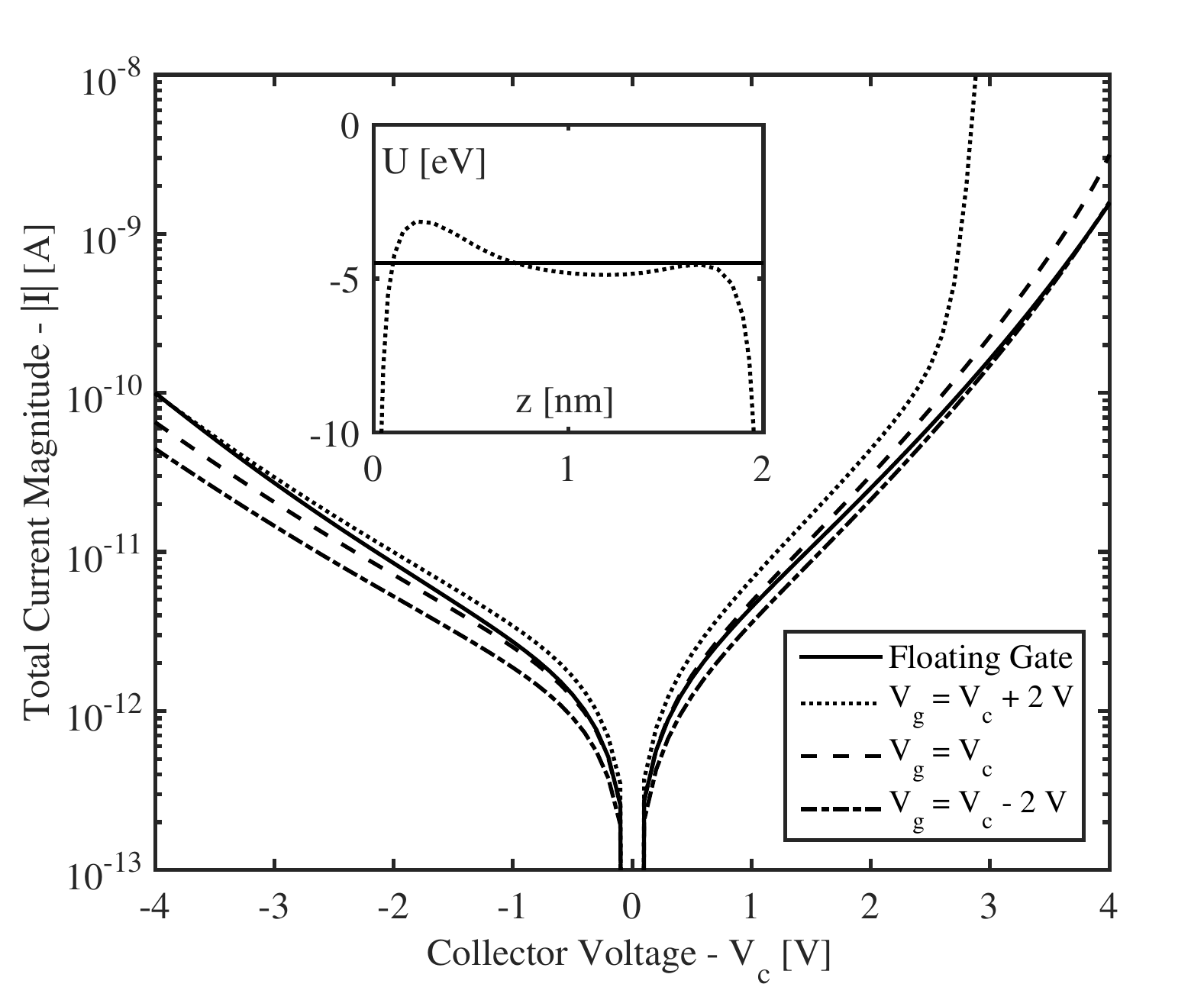}
\par\end{centering}

}\caption{Current magnitude versus collector voltage at different gate connections.
The gate is connected through a DC source of values $2$, $0$ and
$-2\text{ }\mbox{V}$ to the emitter in (a) and to the collector in
(b). The insets show the potential energy along the $z$-direction
at $\rho=2\text{ }\mbox{nm}$, when $V_{c}=-2.5\text{ }\mbox{V}$
and $V_{g}=2\text{ }\mbox{V}$ in (a), and when $V_{c}=2.5\text{ }\mbox{V}$
and $V_{g}=V_{c}+2\text{ }\mbox{V}=4.5\text{ }\mbox{V}$ in (b). The
horizontal solid lines in the insets indicate the Fermi level in the
emitting electrode (the collector in (a) and the emitter in (b)).\label{fig:Collector-and-Emitter}}
\end{figure}

$\FiguresCap$\ref{fig:Emitter-Connection} and \ref{fig:Collector-Connection}
show the magnitudes of the currents versus the collector voltage for
the three values of $V_{DC}$ in cases (a) and (b) of the gate connection,
respectively. The magnitude of the current in case (c), where the
gate potential is floating, is included in both figures as a reference.
The collector voltage is varied from $-4$ to $4\text{ }\mbox{V}$,
which corresponds to applying electric fields of magnitudes between
$0$ and $2\text{ }\mbox{Vnm\ensuremath{{}^{-1}}}$ on the device along
the $z$-direction.

In the case of the floating gate potential, no difference is observed
in both forward and backward currents from the original case, in which
no gate existed. In order to understand the results of the other gate
connections, we refer to the results of the floating gate potential
as the reference results. We also refer to the self-biased gate voltage
in this case, which is $V_{c}/2$, as the gate reference voltage $V_{ref}$.
At $V_{g}$ greater than the reference voltage, the barrier drops
below its reference height (its height when the gate voltage is floating).
So, the current increases such that if $V_{c}$ is positive the forward
current increases, while if it is negative the backward current increases.
At $V_{g}<V_{ref}$, the barrier rises above its reference height
and the current decreases. If the vertical position of the gate is
changed between the emitter and the collector, the reference voltage
will change. For example, if the gate is centered around $3/4$ of
the distance $D$ between the emitter and the collector, then the
gate reference voltage will be $\frac{3}{4}V_{c}$, above which the
current increases and below which the current decreases, keeping its
direction controlled by $V_{c}$ polarity.

Based on this argument the results in $\Figure$\ref{fig:Collector-and-Emitter}
can be explained. In the case when the gate is connected directly
to the emitter ($V_{g}=0$), the gate voltage is lower than the reference
voltage for a positive $V_{c}$, while it is higher than it for a negative
$V_{c}$. Thus, the forward current is lower than the reference forward
current, and the backward current is higher than the reference backward
current. At $V_{g}=-2\text{ }\mbox{V}$, the reference voltage is
greater than the gate voltage as long as $V_{c}$ is greater than
$-4\text{ }\mbox{V}$ (because $V_{ref}=V_{c}/2$), and therefore
the current magnitude is lower than the reference current. Exactly
at $V_{c}=-4\text{ }\mbox{V}$, the two curves intersect. For lower
values of $V_{c}$ the backward current of $V_{g}=-2\text{ }V$ exceeds
the reference backward current. Similarly, at $V_{g}=2\text{ }\mbox{V}$
the current curve intersects with the reference current at $V_{c}=4\text{ }\mbox{V}$.
However, in this case the backward current increases dramatically
at collector voltages below $-2.5\text{ }\mbox{V}$. This is because
the Fermi level in the emitting electrode (the collector in this case)
raised above one of the vacuum barriers at $\rho=2\text{ }\mbox{nm}$
(see the inset in $\Figure$\ref{fig:Emitter-Connection}). Electrons
at the Fermi level, thus, encounter a single narrow barrier, which
increases their tunneling probability. As the collector voltage decreases,
the Fermi level inside the collector increases leading to a narrower
barrier and hence a higher backward current. Such an increase in the
current occurs when the gate potential is large compared to the potential
of the emitting electrode. This effect can be used for switching the
current on and off by varying the gate potential only, ignoring the
rectification effect.

In a similar manner, the current curves when the gate is connected
to the collector through a DC source in $\Figure$\ref{fig:Collector-Connection}
can be understood. In this case, however, all the effects are reversed.
For example, at $V_{g}=V_{c}$ the forward current increases over
its reference value, while the backward current is suppressed. The
potential energy profile in this case is the same as if a single electrode
of applied voltage $V_{c}$ is extended from the collector to the
gate along the outer side of region IV. This structure represents
a diode of inverted U-shaped (or concave) collector. The obtained
current results then imply that such a diode shall have enhanced rectification
properties over the original diode of a flat collector.

At $V_{g}=V_{c}+2\text{ }\mbox{V}$, the forward current increases
dramatically at collector voltages above $2.5\text{ }\mbox{V}$ in
an analogous way to the increase of the backward current at $V_{g}=2\text{ }\mbox{V}$.
This time, however, the increase in the current is in favor of the
rectification behavior of the device. We note that in both cases this
effect appears when the difference between the gate voltage and the
emitting electrode voltage is around $4.5\text{ }\mbox{V}$, which
is the value of the work function of the metal.

\begin{figure}
\begin{centering}
\includegraphics[scale=0.65]{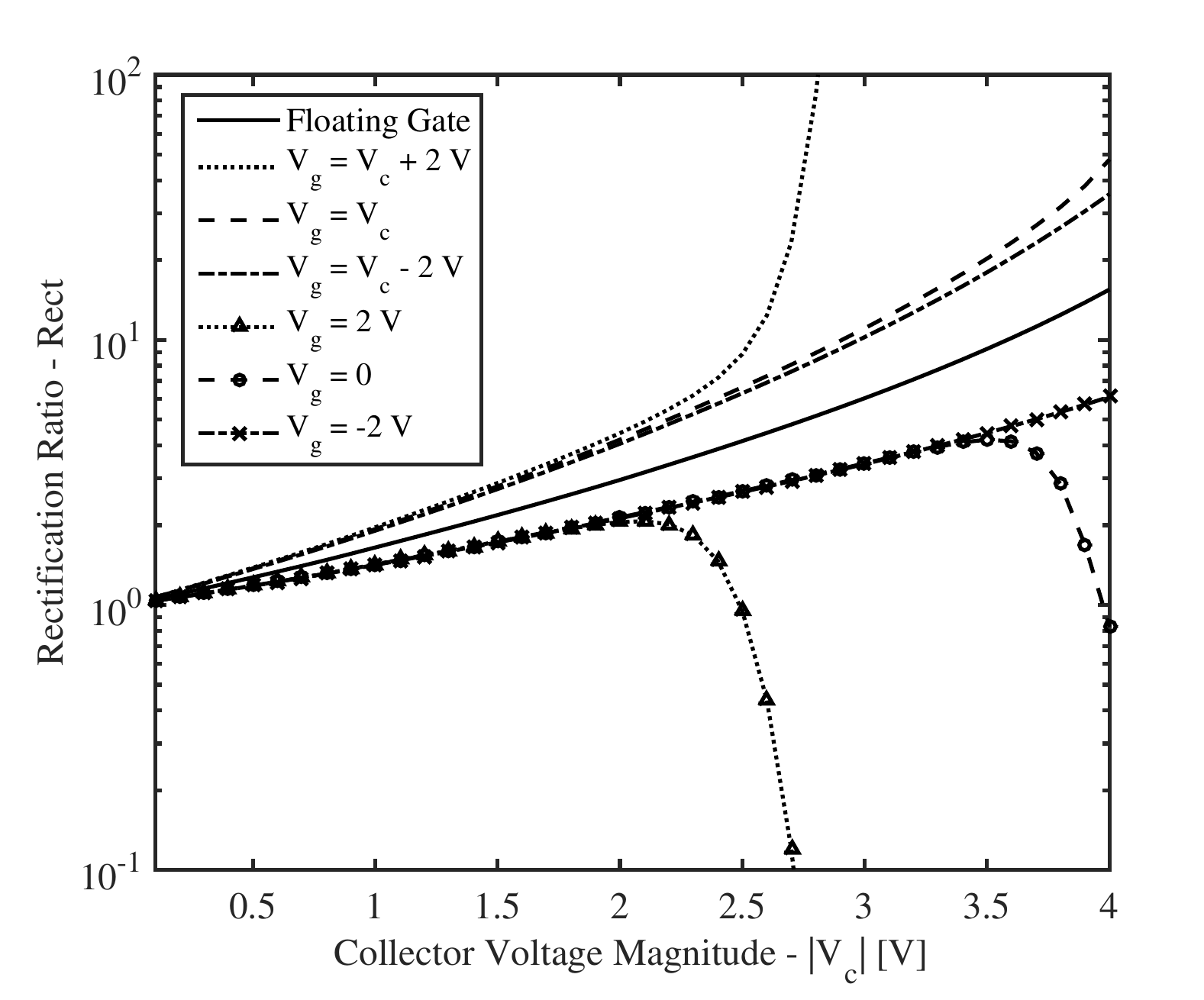}
\par\end{centering}

\caption{Rectification ratio at the different gate connections. The ratio is
calculated for the forward and backward currents in $\protect\Figure\ref{fig:Collector-and-Emitter}$.\label{fig:Rectification-Ratio}}
\end{figure}

\begin{figure}
\centering{}\includegraphics[scale=0.65]{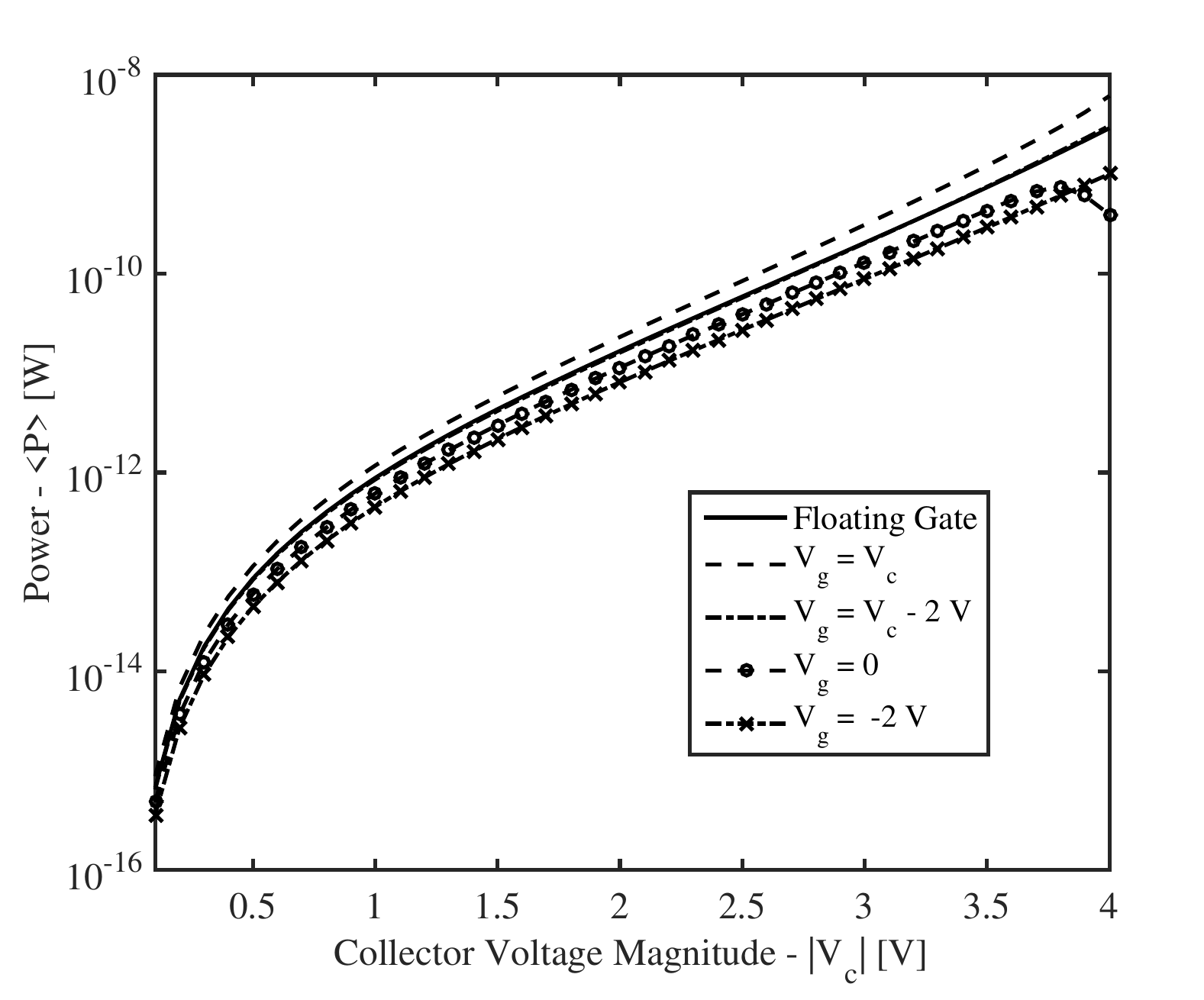}\caption{Output power at the different gate connections. The power is calculated
for the corresponding forward and backward currents in $\protect\Figure\ref{fig:Collector-and-Emitter}$.\label{fig:Output-Power}}
\end{figure}

The rectification ratio and the magnitude of the output power for
the different gate connections are presented in $\Figures$ \ref{fig:Rectification-Ratio}
and \ref{fig:Output-Power}, respectively. The results show that both
the rectification ratio and the output power are enhanced, compared
to the floating gate situation, when the gate is connected directly
to the collector, while they deteriorate when the gate is connected
to the emitter through a DC source.

In the conclusion of this section we can summarize how the gate electrode
can be exploited to enhance the output power of the device. In order
to increase the forward current, the gate voltage should be greater
than the reference voltage, however this may also increase the backward
current even higher and result in decreasing the rectification ratio.
On the other hand, to increase the rectification ratio the gate should
have a DC value referred to the collector not the emitter. This means
that it should be connected to the collector with DC shift, that is
$V_{g}=V_{c}+V_{DC}$ with respect to the emitter. Combining the previous
two results, the output power can be enhanced by connecting the gate
to the collector through DC source whose value is larger than negative
the difference between $V_{c}$ and the reference voltage ($V_{DC}>-(V_{c}-V_{ref})$).
As $V_{DC}$ increases, forward current, rectification ratio and the
mean output power will increase. However, it should not increase much
above zero to avoid resonant tunneling.

There is one remaining note regarding the validity of the model assumptions
in power calculations. Since the leakage current in the gate is neglected,
there is no power consumed or delivered through the gate electrode.
In practice, however, this approximation is not very accurate, particularly
in the case of high gate voltages, where any small current leakage
in the gate would result in considerable power exchange through it. If the
device is to be used in energy conversion applications, gate current
should be included in power calculations, otherwise the calculated
power efficiency of the device would be misleading.

\subsection{Dependence of the Collector Current Modulation on the Gate Parameters\label{sub:Current-Modulation}}

In this section, we investigate the ability of the nanotriode to modulate
the collector current using the gate voltage, and the modulation dependence
on the geometrical parameters of the gate. The collector voltage is
fixed at $1\text{ }\mbox{V}$, and only the forward current is considered
in the calculations. We take values of $V_{g}$ between $-2$ and
$2\text{ }\mbox{V}$. All the geometrical parameters are similar to
$\Section$\ref{sub:Barrier-Modulation} except that the gate aperture
diameter is $4.2\text{ }\mbox{nm}$. This diameter is big enough to
avoid resonant tunneling at the maximum gate voltage value used in
this section ($2\text{ }\mbox{V}$).

\begin{figure}
\begin{centering}
\includegraphics[scale=0.65]{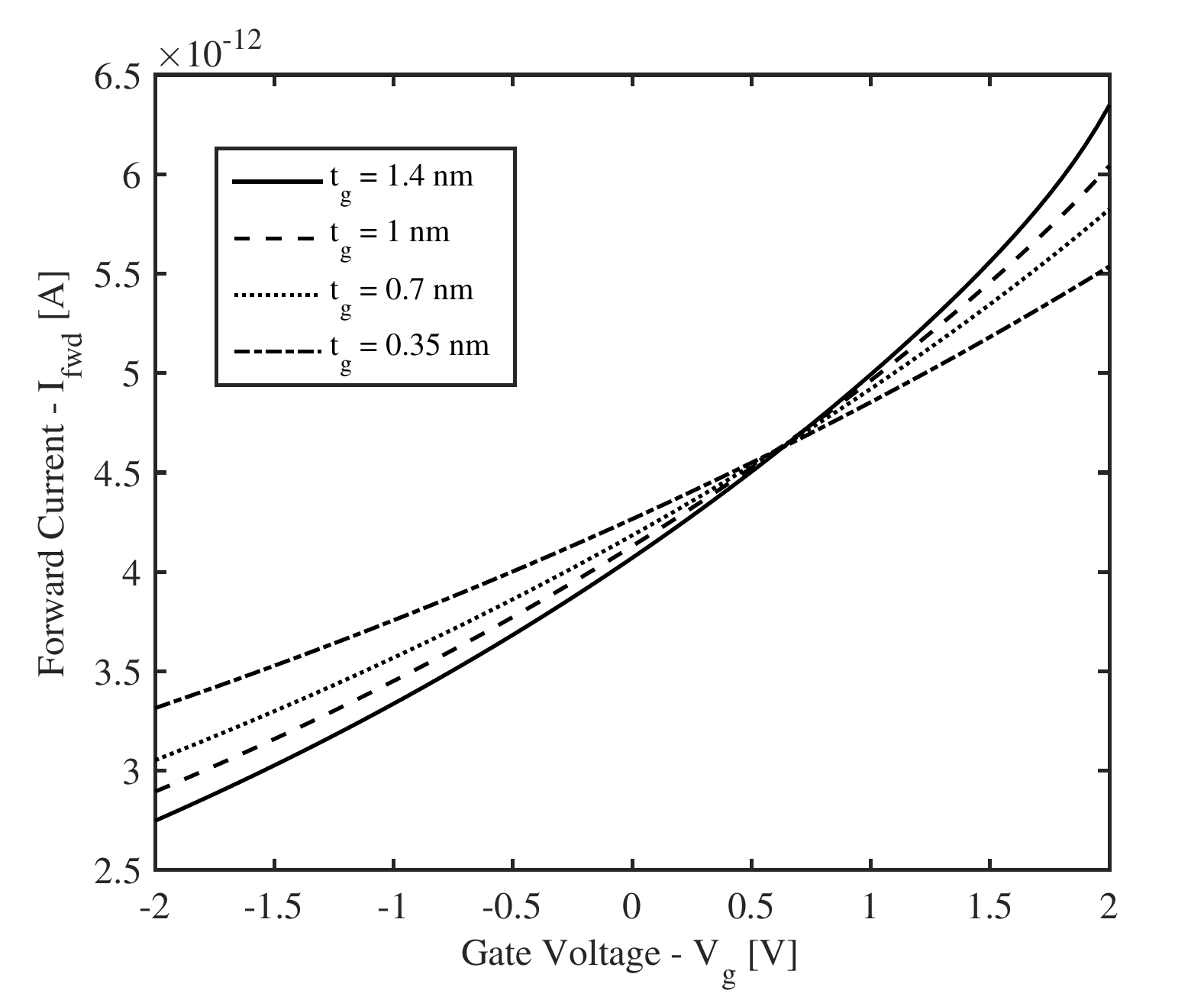}
\par\end{centering}

\caption{Forward current versus gate voltage at collector voltage $V_{c}$
of $1\text{ }\mbox{V}$ for four different values of gate thickness;
$t_{g}=1.4$, $1$, $0.7$ and $0.35\text{ }\mbox{nm}$. The gate
height $h_{g}$ and aperture diameter $d_{g}$ are $1$ and $4.2\text{ }\mbox{nm}$,
respectively. \label{fig:Gate-Thickness-Effect}}
\end{figure}

We start with the effect of the gate thickness $t_{g}$ on the triode's
ability to alter the current through the gate voltage $V_{g}$. $\FigureCap$\ref{fig:Gate-Thickness-Effect}
shows the forward current as a function of the gate voltage at four
different gate thicknesses. Since we are interested in studying the
nanotriode in its smallest practical dimensions, we consider the
smallest possible thickness for a monolayer conducting sheet. This
is estimated to be in the range of $0.35\text{ }\mbox{nm}$ for a
graphene layer \cite{Lu1997,Jussila2016}. On the upper limit, we
are restricted by the separation between the collector and the emitter.
A gate with thickness approaching this separation would result in
high-probability tunneling between the emitter and the collector through
the gate, turning the device into a conductor. Structures with non-flat
gate, such as tapered gate, around the tip will be studied in future
work. $\FigureCap$\ref{fig:Gate-Thickness-Effect} shows that as
the gate thickness increases, the variation of the current with the
gate voltage increases. At a gate thickness of $1.4\text{ }\mbox{nm}$,
an increase in the gate voltage from $-2$ to $2\text{ }\mbox{V}$
causes an increase in the current by $130\%$.

\begin{figure}
\begin{centering}
\includegraphics[scale=0.65]{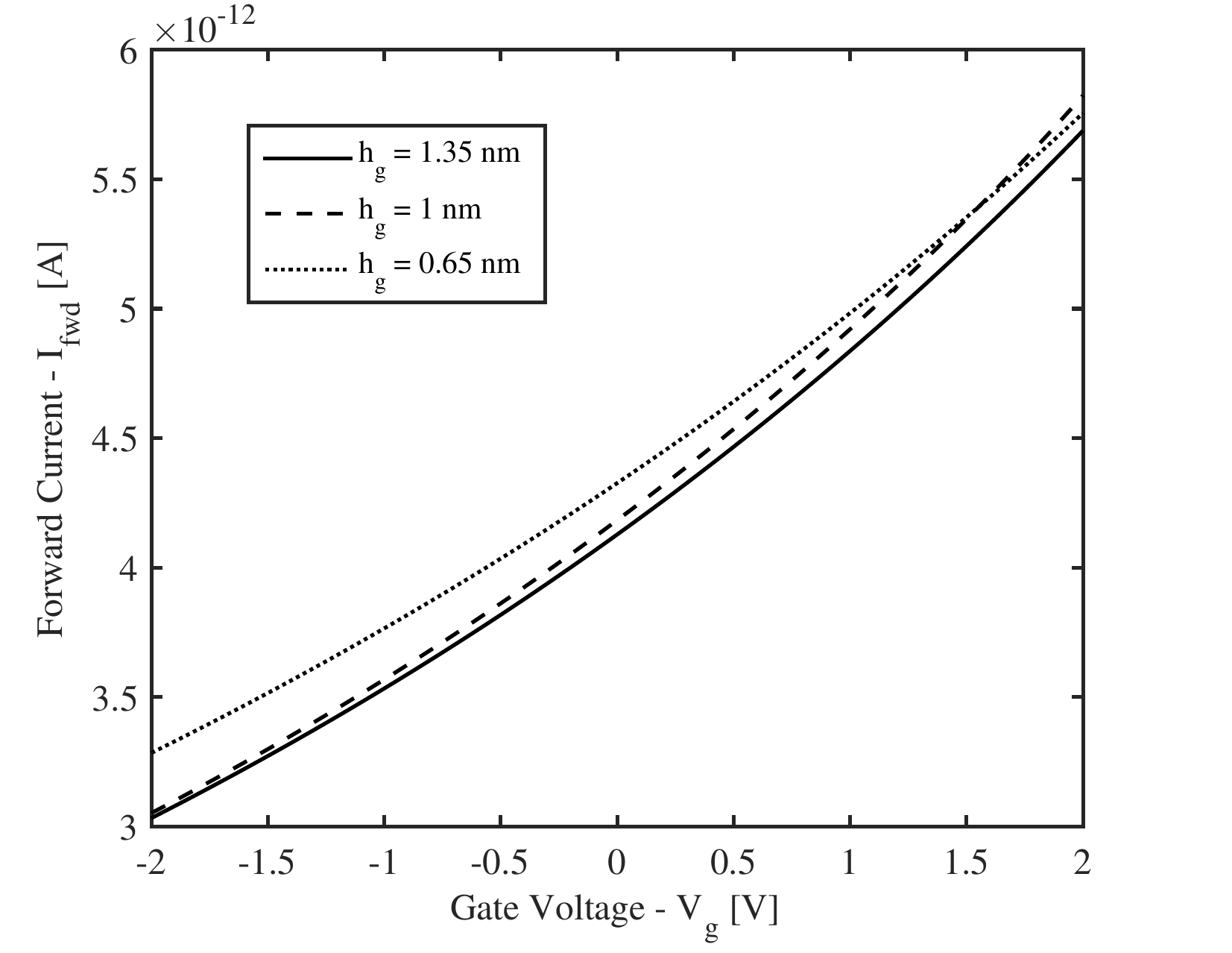}
\par\end{centering}

\caption{Forward current versus gate voltage at collector voltage $V_{c}$
of $1\text{ }\mbox{V}$ and gate thickness $t_{g}$ of $0.7\text{ }\mbox{nm}$
for three different heights of the gate; $h_{g}=1.35$, $1$ and $0.65\text{ }\mbox{nm}$.
The gate aperture diameter $d_{g}$ is $4.2\text{ }\mbox{nm}$. The
chosen three heights correspond, descendingly, to the positions of
the gate where it is just above, centered around, and just below the
end of the tip.\label{fig:Gate-Height-Effect}}
\end{figure}

Turning to the effect of the vertical position of the gate, we present
the current results at gate thickness of $0.7\text{ }\mbox{nm}$ for
three different gate positions in $\Figure$\ref{fig:Gate-Height-Effect}.
The three positions are chosen such that a) the gate is just above
the tip end ($h_{g}=1.35\text{ }\mbox{nm}$), b) the gate is centered
around the tip end ($h_{g}=1\text{ }\mbox{nm}$), and c) the gate
is surrounding the highest part of the tip ($h_{g}=0.65\text{ }\mbox{nm}$).
It is obvious from the results that the largest variation in the current
is obtained when the gate is centered around the end of the tip. That
is because the highest part in the potential barrier at any radial
position within the tip range is just above its curved surface, so
the gate is most effective when centered around this region. The results
in $\Figure$\ref{fig:Gate-Height-Effect} also show that the current
variation is minimal in the third case when $h_{g}=0.65\text{ }\mbox{nm}$.
That is because the gate is the furthest from the barrier between
the end of the tip and the collector surface at the center of the
device ($\rho=0$) where the tunneling current density is the maximum.

\begin{figure}
\begin{centering}
\includegraphics[scale=0.65]{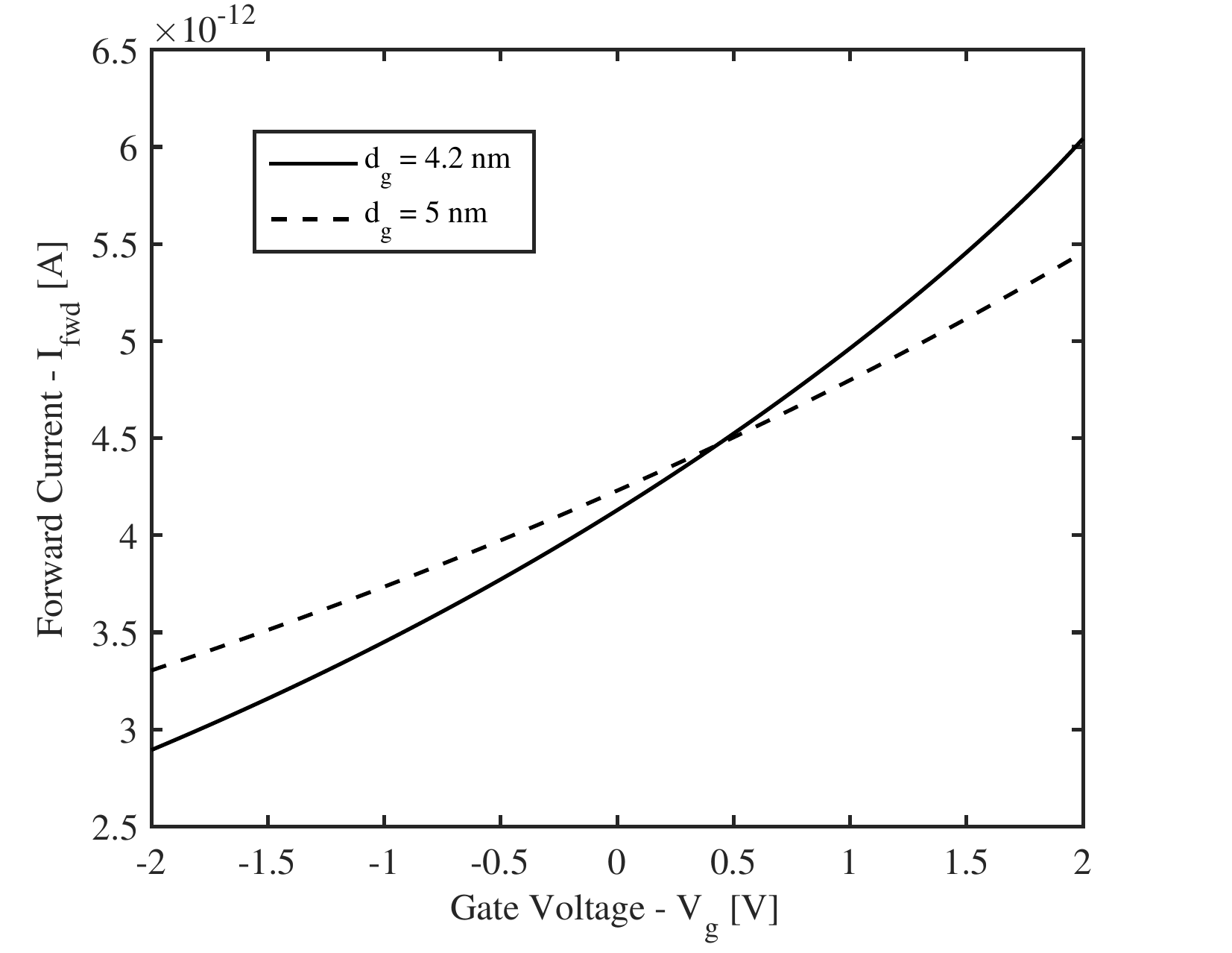}
\par\end{centering}

\caption{Forward current versus gate voltage at collector voltage $V_{g}$
of $1\text{ }V$ and gate thickness $t_{g}$ of $1\text{ }\mbox{nm}$
for two different gate aperture diameters; $4.2$ and $5\text{ }\mbox{nm}$.
The height of the gate is $1\text{ }\mbox{nm}$.\label{fig:Aperture-Diameter-Effect}}
\end{figure}

The effect of the gate aperture diameter on the shape of the potential
barrier is discussed in $\Section$\ref{sub:Barrier-Modulation}.
Now we investigate how this dependency will affect the variation of
the current with varying the gate voltage. $\FigureCap$\ref{fig:Aperture-Diameter-Effect}
shows the $I-V_{g}$ relation for gates of aperture diameters
$4.2$ and $5\text{ }\mbox{nm}$. The gates considered here have a thickness
and a height of $1\text{ }\mbox{nm}$. The decrease of the current variation
with increasing the aperture diameter is clear from the graph. This
result is quite predictable from the potential energy results in $\Figures$\ref{fig:Potential-Barrier-1}
and \ref{fig:Potential-Barrier-2}.

\subsection{Dependence of Current Modulation on the Tip Parameters}

\begin{figure}
\begin{centering}
\includegraphics[scale=0.65]{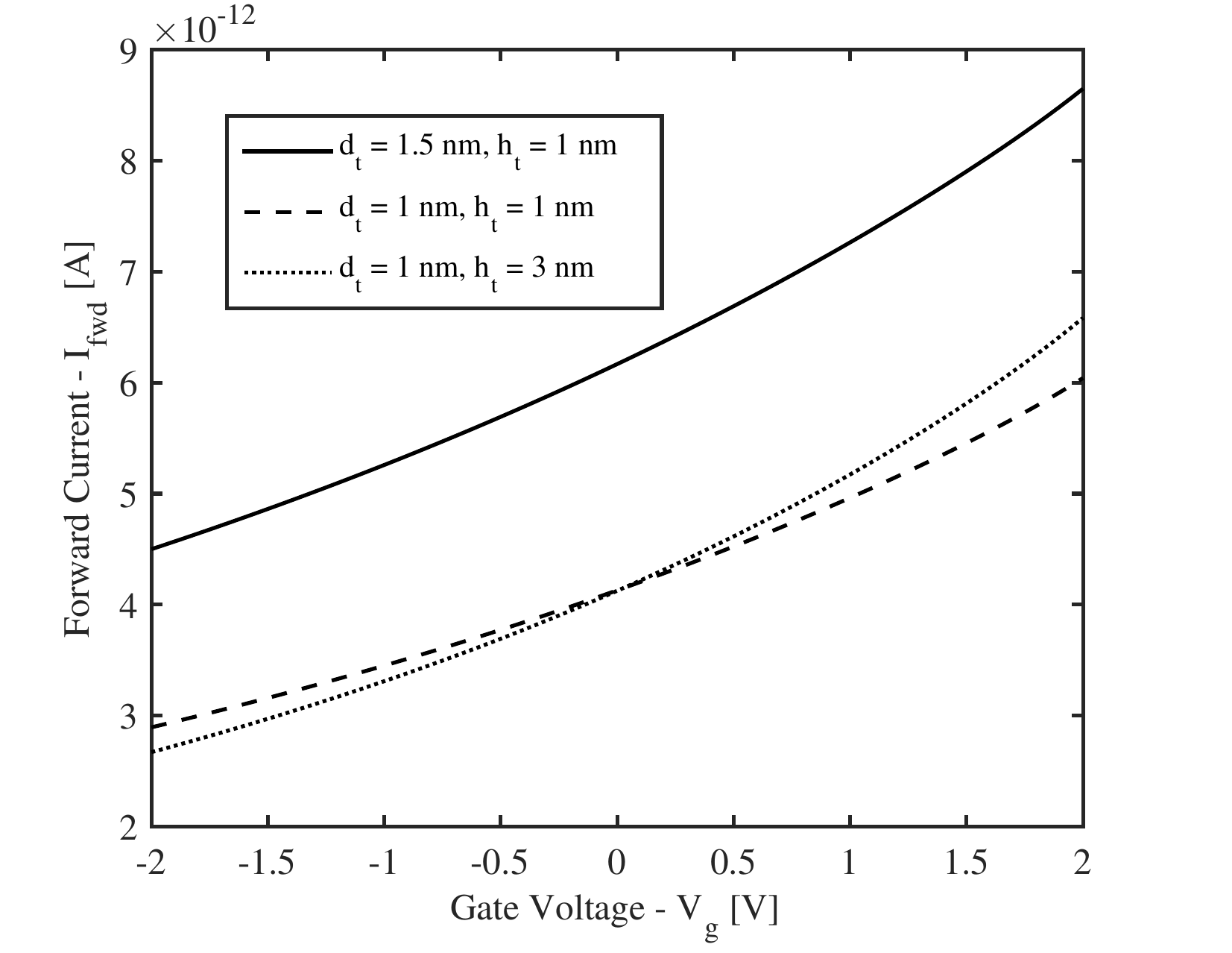}
\par\end{centering}

\caption{Forward current versus gate voltage at collector voltage $V_{c}$
of $1\text{ }\mbox{V}$ for different tip parameters. At the two cases
where $h_{t}=1\text{ }\mbox{nm}$, the separation distance between
the leads $D$ is $2\text{ }\mbox{nm}$. At $h_{t}=3\text{ }\mbox{nm}$,
$D=3\text{ }\mbox{nm}$. In the three cases, the gate thickness $t_{g}$
and aperture diameter $d_{g}$ are $1$ and $4.2\text{ }\mbox{nm}$,
respectively, and $h_{g}=h_{t}$. \label{fig:Tip-Effect}}
\end{figure}

In $\References$\onlinecite{Mayer2008,Mayer2010a}, Mayer et al.
demonstrated that better rectification properties are for emitting
tips with higher aspect ratio. With adding the gate electrode to the
structure, we aim in this section at investigating the effect of the
aspect ratio on current modulation. In this section we use gate thickness,
height and aperture diameter of $1$, $1$ and $4.2\text{ }\mbox{nm}$,
respectively. As shown in $\Figure$\ref{fig:Tip-Effect}, by increasing
the tip diameter $d_{t}$ from $1$ to $1.5\text{ }\mbox{nm}$ at the same tip height ($h_{t}=1\text{ }\mbox{nm}$), an
increase in both the average current and the slope of the $I-V_{g}$
curve is observed. The increase in the average current value is due
to the increase of the emitting area represented by the tip surface.
Such an increase in the forward current is necessarily accompanied
by a higher increase in the backward current, due to the decreasing
field enhancement, resulting at the end in reducing the rectification
ratio. The increase in the slope is mainly because the emitting tip
extended to a closer region to the gate, where the modulating effect
is more significant on the potential energy and the emitted current.

Finally, we investigate the gate modulation effect on tips of different
heights. A tip of height $3\text{ }\mbox{nm}$ and diameter $1\text{ }\mbox{nm}$
is examined. The separation between the leads is $4\text{ }\mbox{nm}$,
so that the width of the potential barrier between the tip and the
collector is $1\text{ }\mbox{nm}$ as in the previous cases. Also
the gate is set at height of $3\text{ }\mbox{nm}$, in order to be
centered around the end of the tip as well. A better current modulation
is observed at this higher tip (dotted curve in $\Figure$\ref{fig:Tip-Effect}).
This result exhibits a favorable behavior for the device due to its
compatibility with the results obtained in $\References$\onlinecite{Mayer2008,Mayer2010a},
where optimized rectification properties were also demonstrated for
higher tips.

\section{Conclusion}

In this paper, a transfer matrix method is used to model quantum tunneling
through vacuum nanotriodes at low applied voltages. The structure
consists of a metal-vacuum-metal junction with a nanotip supported
on one of the metals and surrounded by a thin gate electrode. The
device resembles Spindt-type vacuum triodes with a collector-emitter
separation and a gate aperture diameter of a few nanometers. The details
of calculating the potential energy distribution and the field emission
current are presented with considering the gate effect. The behavior
of the device is then investigated as a current rectifier and modulator
at different electric and geometric parameters.

The results show a significant enhancement in the rectification properties
of the geometrically asymmetric metal-vacuum-metal diode when a gate
electrode is connected to the collector through a DC source. This
finding suggests a better rectification for vacuum nanodiodes with
concave collectors instead of flat ones. It is also demonstrated that
the output current of the device can be modulated by the gate voltage.
The effect of the geometrical parameters of the gate and the tip on
current modulation are investigated. It is shown that a gate centered
around the end of the tip with a narrow aperture and a large thickness
gives the best current modulation. It is also shown that the currents
emitted from tips of higher aspect ratios are better controlled
through the gate potential. In order to achieve electric currents
of higher magnitudes at the same applied potentials, we have to either
decrease the gap distance between the emitting tip and the collecting
surface, or increase the emitting surface area. The second solution
can be realized by adding more than one nanotip or a protruding ring.
Such structures can be designed and characterized following the same
approach we present in this work. However, in order for this model
to be fully reliable in the design and the characterization processes,
experimental verification on similar dimensions is needed.

Based on this analysis, the vacuum nanotriode we study in this paper
represents an excellent candidate for high-frequency applications.
It can be optimized to do the basic functions of transistors, in addition to its rectification
effect, at high
frequencies up to the infrared range. Thus, it is a possible alternative to semiconductor transistors
as a basic unit in electronic circuits. It can also be used for implementing
local processing operations on rectennas, avoiding the transmission
problems of high-frequency signals. This shall open the door to a
new era of fast electronics, where communication systems and processing
units operating at optical frequencies are achievable. It can also
be used as an electronic interface for plasmonic circuits.
\begin{acknowledgments}
M. Khalifa would like to thank A. Mayer for his valuable comments and suggestions.
\end{acknowledgments}

\bibliography{References}

\end{document}